\documentclass[preprint,review,12pt]{revtex4}
\usepackage{amsmath}
\usepackage{color}
\usepackage[dvips]{graphicx}
\newcommand{\ep}{\varepsilon}
\begin{document}
\title{Appearance of Thermal Time}

\author{Shigenori Tanaka}
\email{tanaka2@kobe-u.ac.jp}
\affiliation{Graduate School of System Informatics, Kobe University, 1-1 Rokkodai, Nada, Kobe 657-8501, Japan}

\date{\today}

\begin{abstract}

In this paper a viewpoint that time is an informational and thermal entity is shown. 
We consider a model for a simple relaxation process for which a relationship among event, time and temperature is mathematically formulated. 
It is then explicitly illustrated that temperature and time are statistically inferred through measurement of events. 
The probability distribution of the events thus provides a mutual regulation between temperature and time, which can relevantly be expressed in terms of 
the Fisher information metric. 
The two-dimensional differential geometry of temperature and time then leads us to a finding of a simple equation for the scalar curvature, $R = -1$, in this case of relaxation process.
This basic equation, in turn, may be regarded as characterizing the nonequilibrium dynamical process and having a solution given by the Fisher information metric.
The time can then be interpreted so as to appear in a thermal way. 

\vspace{1.0cm}
\\

\end{abstract}

\maketitle

\setlength{\baselineskip}{24pt}

\section{Introduction}

In earlier investigations, there have been some suggestions and proposals that the time may emerge in thermal way \cite{Rovelli1,Rovelli2}. 
Among a lot of discussions on the emergence of time \cite{Davies,Ellis}, there is a novel viewpoint that the time originates as a statistical or thermodynamic entity. 
The main purpose of this paper is to illustrate a simple model to embody this notion.
The key idea is as follows. 
We emprically recognize that the temperature is a coarse-grained, macroscopic parameter that is estimated statistically through measurements on some phenomena in our world.
Then, instead of presuming that the time acts as an independent, {\it a priori} variable, we regard it as a coarse-grained parameter inferred {\it a posteriori} as well as the temperature.
To this end, we consider a nonequilibrium relaxation process for which the time and temperature parameters are analyzed in the framework of information geometry 
associated with some statistical events.  

\par

In the present modeling, we consider a phenomenon represented by a dynamical variable $x$ that accords to a simple stochastic relaxation process. 
We especially focus on a linear relaxation dynamics of $x$ in harmonic potential at temperature $T$, which is known as the Onsager-Machlup (OM) process \cite{Onsager}. 
Assuming an intial condition that $x$ is localized at a point $x_0$, we can solve the Fokker-Planck equation for this relaxation process 
to find an analytic expression \cite{Chandrasekhar,Yasue,Tanaka2} for time-dependent probability 
distribution function $P(x,t)$.
Then, we rewrite this distribution function as $P(x;t,\beta)$ or $P(\{x\}|t,\beta)$ with $\beta=1/k_{B}T$ ($k_{B}$ is the Boltzmann constant) 
to represent that the distribution of $x$ is described given two parameters, time $t$ and (inverse) temperature $\beta$.
Here, we know that the temperature is a coarse-grained parameter determined through measurements on the statistical behavior of the variable $x$, and 
we further presume that the time may also be the case.
In the terminology of information theory \cite{MacKay}, $P(\{x\}|t,\beta)$ expresses the likelihood of two parameters $t$ and $\beta$ for given data set of $\{x\}$. 
The Fisher information metric \cite{Frieden,Amari} for $t$ and $\beta$ can then give a geometrical relationship between the two parameters in terms of the statistical distribution or measurement of $x$.
According to the two-dimensional differential geometry in $(\beta,t)$ space, we will be led to some tensor expressions to describe the interrelated behaviors of $t$ and $\beta$.
We thus find in this case a simple differential equation represented by the scalar curvature, $R = -1$, which we regard as a basic equation to govern the two coarse-grained parameters $\beta$ and $t$ 
associated with the OM relaxation process. 

\par 

In the following Sec.\ II, we first illustrate how two coarse-grained parameters, temperature and time, can be statistically estimated 
in a linear relaxation process of the OM type. 
In Sec.\ III, the Fisher information metric for time and temperature is derived on the basis of the OM relaxation process. 
Furthermore, after some algebra in differential geometry, we are led to a simple expression for the scalar curvature to characterize the two-dimensional information space of time and temperature;  
the mathematical details for the derivation are found in the author's earlier paper \cite{Tanaka2}. 
In Sec.\ IV, the logical way is reversed so that one can see the mutual relation between time and temperature as a solution to the basic equation for the scalar curvature.
Some concluding remarks are given in Sec.\ V.

\section{Statistical Inference of Temperature and Time in the Onsager-Machlup Process}

As a model system, this study considers a temporal relaxation process of dynamical variable $x$ described by an overdamped Langevin equation 
at temperature $T$ \cite{Nitzan,Sekimoto,Faccioli,Sega,Adib,Tanaka}, 

\begin{equation}
\dot{x}=-\frac{D}{k_{B}T}U'(x)+\eta(t),
\end{equation}

\noindent
where $k_{B}$, $D$ and $U'(x)$ refer to the Boltzmann constant, the diffusion coefficient 
and the derivative of potential energy, respectively. 
$\eta(t)$ is a Gaussian noise with zero average satisfying the fluctuation-dissipation relation, 

\begin{equation}
\langle \eta(t)\eta(t')\rangle = 2D\delta(t-t'),
\end{equation}

\noindent
where $\langle \quad \rangle$ means the statistical average. 
The time-dependent probability distribution function $P(x,t)$ sampled by the stochastic differential equation (1) then obeys 
the Fokker-Planck-Smoluchowski equation \cite{Nitzan,Sekimoto,Adib}, 

\begin{equation}
\frac{\partial}{\partial t}P(x,t) = D\frac{\partial}{\partial x}\left[\frac{\partial}{\partial x}P(x,t) + \beta U'(x)P(x,t)\right] 
\end{equation}

\noindent
with $\beta=1/k_{B}T$.

\par

In this study we focus on a (linear) dynamics under the harmonic potential, 

\begin{equation}
U(x) = \frac{1}{2}kx^{2},
\end{equation}

\noindent
with the spring constant $k$. 
Then, assuming a localized initial distribution, 
$P(x,0) = \delta(x-x_{0}) \ (x_{0} \ne 0)$, a relaxation process of the Onsager-Machlup (OM) type \cite{Onsager} is realized. 
With $\gamma = \beta kD$, an explicit expression for 
the time-dependent probability distribution function is found to be \cite{Yasue,Tanaka2} 

\begin{equation}
P(x,t) = \sqrt{\frac{\beta k}{2\pi(1-e^{-2\gamma t})}}\exp\left[-\frac{\beta k(x-x_{0}e^{-\gamma t})^{2}}{2(1-e^{-2\gamma t})}\right], 
\end{equation}

\noindent
which is a well-known expression \cite{Chandrasekhar}. 
Here, we see that the statistical distribution of ``phenomenon" $x$ is governed by the time and temperature parameters $t$ and $\beta$, along with model specific parameters $k$, $D$ and $x_0$. 
Alternatively, Eq.\ (5) mathematically represents a dynamical model structure that our world has approximately. 
To imply this viewpoint explicitly, we express the distribution function as $P(x;t,\beta)$ or $P(\{x\}|t,\beta)$. 
In the limit of $t \to \infty$, $P(x;t,\beta)$ approaches the Boltzmann distribution with the harmonic potential $U(x)$. 

\par

Now, let us consider a problem of how we can infer the temperature and time when we observe the phenomena $x$ whose dynamics is described in terms of Eq.\ (5) in our world. 
Given the conditional probability distribution function $P(x|\lambda)$ with the parameter $\lambda$, when we have obtained the data $\{x_i\}\ (i=1,2,...)$ for the quantity $x$, 
we can statistically infer the parameter on the basis of Bayes' theorem as \cite{MacKay} 

\begin{equation}
P(\lambda|\{x_i\}) = \frac{P(\{x_i\}|\lambda)P(\lambda)}{P(\{x_i\})}, 
\end{equation}

\noindent
where $P(\lambda)$ and $P(\{x_i\})$ refer to the prior probability for $\lambda$ and the evidence for $\{x_i\}$, respectively. 
If we presume these probabilities are constant {\it a priori}, we will find 

\begin{equation}
P(\lambda|\{x_i\}) \propto P(\{x_i\}|\lambda) = \prod_{i}P(x_i|\lambda).  
\end{equation}

\par

Here, we set $k, D, x_{0} = 1$ for simplicity, and assume that we have obtained the data $\{x_i\} = \{0.25,0.30,0.35,0.37,0.40\}$ at $t = 0.5$, for example. 
Employing Eq.\ (5), we can then depict the right-hand side of Eq.\ (7) as a function of $\beta$ in Fig.\ 1(a).
We can thus know, through the measurement of $\{x_i\}$, the most probable value of (inverse) temperature as $\beta = 5.8$ at which the probability or the likelihood is maximal. 
This may be a way in which we observe the temperature statistically. 
On the other hand, if we have obtained the data $\{x_i\} = \{0.25,0.30,0.35,0.37,0.40\}$ at $\beta = 5.0$, for example, 
we can depict the right-hand side of Eq.\ (7) as a function of $t$ in Fig.\ 1(b).
In this case we obtain an inference that the most probable value of time is $t = 0.18$ through the measurement. 
If we collect much more information on the data $\{x_i\}$, we will find a sharper peak in the likelihood $P(\{x_i\}|\lambda)$.

\section{Information Geometry of Relaxation Process}

In the preceding section, we have seen that the most probable values of $\beta$ and $t$ can be inferred from the data $\{x_i\}$ 
when we assume the relaxation process of the OM type.
In this section, we mathematically address the geometry of the parameter $(\beta,t)$ space which regulates the probability distribution of the phenomena $x$.

\par

The Fisher information metric is a particular Riemannian metric in the space of probability distributions and plays a central role in information geometry \cite{Frieden,Amari}. 
It can be used to calculate the informational difference between measurements as the infinitesimal form of the relative entropy such as the Kullback-Leibler divergence \cite{MacKay}.
In the present study based on the probability distribution function $P(x)$ for the event $\{x\}$ , the Fisher information metric is given by 

\begin{eqnarray}
g_{\mu\nu} &=& \langle(\partial_{\mu}\ln P)(\partial_{\nu}\ln P)\rangle \nonumber \\
 &=& \int dx P(x)\left[\partial_{\mu}\ln P(x)\right]\left[\partial_{\nu}\ln P(x)\right] 
\end{eqnarray}

\noindent
in the parameter space represented by $\mu$ and $\nu$.
In particular, we consider in this study the two-dimensional parameter space formed by the inverse temperature 
$\beta = 1/k_{B}T$ and the time $t$ to characterize the nonequilibrium process described by $P(x;t,\beta)$ of Eq.\ (5). 
The (covariant) metric tensor is then calculated to be 

\begin{equation}
g_{\beta\beta} = \frac{1}{2\beta^{2}}\left(1-2\gamma t\frac{\ep}{1-\ep}\right)^{2} + \frac{kx_{0}^{2}}{\beta}\gamma^{2}t^{2}\frac{\ep}{1-\ep}, 
\end{equation}

\begin{equation}
g_{tt} = 2\gamma^{2}\left(\frac{\ep}{1-\ep}\right)^{2} + kx_{0}^{2}\beta\gamma^{2}\frac{\ep}{1-\ep}, 
\end{equation}

\begin{equation}
g_{\beta t} = g_{t\beta} = -\frac{\gamma}{\beta}\frac{\ep}{1-\ep}\left(1-2\gamma t\frac{\ep}{1-\ep}\right) + kx_{0}^{2}\gamma^{2}t\frac{\ep}{1-\ep}, 
\end{equation}

\noindent
with the determinant, 

\begin{equation}
g = \det(g_{\mu\nu}) = \frac{kx_{0}^{2}\gamma^{2}\ep}{2\beta(1-\ep)}, 
\end{equation}

\noindent
where $\ep = e^{-2\gamma t} = e^{-2\beta kDt}$ has been introduced. 
The contravariant metric tensor, {\it i.e.}, the inverse matrix of the Fisher information metric, is accordingly obtained as  

\begin{equation}
g^{\beta\beta} = g_{tt}/g, 
\end{equation}

\begin{equation}
g^{tt} = g_{\beta\beta}/g, 
\end{equation}

\begin{equation}
g^{\beta t} = g^{t\beta} = -g_{\beta t}/g = -g_{t\beta}/g. 
\end{equation}

\par

Then, according to standard algebra in differential geometry \cite{Landau}, the Ricci tensor $R_{\mu\nu}$ is found to satisfy a relation, 

\begin{equation}
R_{\mu\nu} = -\frac{1}{2}g_{\mu\nu}, 
\end{equation}

\noindent
in the present case \cite{Tanaka2}. 
This leads to the scalar curvature given by 

\begin{equation}
R = g^{\mu\nu}R_{\mu\nu} = -\frac{1}{2}g^{\mu\nu}g_{\mu\nu} = -1 
\end{equation}

\noindent
in two dimension. 
The Einstein tensor is then 

\begin{equation}
G_{\mu\nu} = R_{\mu\nu}-\frac{1}{2}Rg_{\mu\nu} = 0, 
\end{equation}

\noindent
which should be the case due to the symmetries in two dimension \cite{Brown}. 

\par

Thus, we have found that the Fisher information metric given by Eqs.\ (9)-(11), which was derived from the $P(x;t,\beta)$ for the OM process, gives $R = -1$ 
in the two-dimensional differential geometry for the parameter space of $\beta$ and $t$, irrespective of the values of $D$, $k$ and $x_{0}$. 
We may regard this as a characterization for the geometry formed by the two parameters of $\beta$ and $t$ \cite{Tanaka2}.

\section{Basic Equation for Thermal Time and Its Solution}

Here, let us review the analysis illustrated in the preceding sections. 
We start with Eq.\ (5) to describe the temporal ($t$) evolution of the probability density function $P(x;t,\beta)$ for a variable $x$ with temperature parameter $\beta$ and specific model 
parameters $k, D$ and $x_0$ for a nonequilibrium relaxation (Onsager-Machlup) process. 
We may then interpret this equation as follows: 
Through measurement of phenomenon $x$ in our world, whose behavior is mathematically expressed by Eq.\ (5), we can empirically, experimentally or statistically estimate 
the two parameters, time $t$ and (inverse) temperature $\beta$, as shown in Sec.\ II. 
The Fisher information metric expressed as Eqs.\ (9)-(11) then gives the regulatory relationship between $t$ and $\beta$ in the two-dimensional parameter space, along with 
specific model parameters $k, D$ and $x_0$ to describe the Onsager-Machlup (OM) process. 
Given the covariant metric tensors in the $(t,\beta)$ space, we can calculate a variety of associated tensors in the two-dimensional differential geometry, 
thus leading to a simple equation for the scalar curvature, $R = -1$, irrespective of the specific model parameters.

\par

Now, this logic can be reversed.
That is, we can regard this simple relation, $R = -1$, as a basic equation to regulate the two variables $t$ and $\beta$ involved in the OM process.
In general, the differential geometry in two dimension can be characterized by only one degree of freedom \cite{Brown}, and therefore the equation $R = -1$ can be regarded as a basic equation 
like the Einstein equation for the Einstein tensors in the case of four-dimensional temporal-spatial geometry. 
Then, we know a solution to this equation under relevant boundary condition for specific physical model, that is Eqs.\ (9)-(11). 
Thus, starting with the ``basic" equation for the OM process, Eq.\ (17), we may find the geometry of the parameter (information) space of $(t,\beta)$, as expressed by the Fisher information metric. 

\par

Then, let us see the behaviors of the information metric $g_{\mu\nu}$. 
In the following we set $k, D, x_{0} = 1$ for illustration. 
Figure 2 illustrates the behaviors of $g_{\beta\beta}$ as functions of $t$ and $\beta$ for various values of $\beta$ and $t$, respectively. 
For $t \to 0$, $g_{\beta\beta}$ vanishes as $t/2$ as shown in Fig.\ 2(a), 
which means that the temperature cannot be defined when the ``particle" or ``event" $x$ is localized at one point $x_0$ in the initial state.
On the other hand, when considering the limit of $t \to \infty$, we find $g_{\beta\beta} \to 1/2\beta^2$, thus corresponding to the thermal equilibrium state in which the temperature is well defined. 
When looking at the $\beta$ dependence of $g_{\beta\beta}$ in Fig.\ 2(b), we observe that $g_{\beta\beta}$ goes to $1/2\beta^2$ and $(t^2 + t)/2$ in the 
low-temperature ($\beta \to \infty$) and high-temperature ($\beta \to 0$) limits, respectively. 

\par

Next, Fig.\ 3 shows the behaviors of $g_{tt}$ as functions of $t$ and $\beta$ for various values of $\beta$ and $t$, respectively.
As shown in Fig.\ 3(a), $g_{tt}$ vanishes in the limit of $t \to \infty$, implying that the time loses its sense in the thermodynamic equilibrium state, 
while $g_{tt} \to 1/2t^2$ for $t \to 0$. 
Concerning the $\beta$ dependence illustrated in Fig.\ 3(b), we see $g_{tt} \to 0$ 
in the low-temperature limit of $\beta \to \infty$, thus implying that the time does not exist because all the degrees of freedom for motion are frozen at zero temperature in classical mechanics.
In the high-temperature region of $\beta \to 0$, on the other hand, we see $g_{tt} \to 1/2t^2$, showing what we expect for the behavior of usual ``time". 
Finally, Fig.\ 4 shows the behaviors of $g_{\beta t}$, in which we observe that $g_{\beta t}$ vanishes in the limit of $t \to \infty$ or $\beta \to \infty$, 
thus indicating the decoupling between time and temperature. 
In this way, we see how the time emerges thermodynamically in the present model.

\section{Conclusion}

In this study we have discussed the thermal origin of time on the basis of mathematical framework of information geometry. 
We have modelled a nonequilibrium relaxation process of the OM type in terms of a probability distribution function of events $\{x\}$, $P(x;t,\beta)$, 
in which time $t$ and (inverse) temperature $\beta$ are regarded as coarse-grained parameters. 
Relying on Bayes' theorem, temperature and time are inferred statistically through measurements on the events. 
The relationship between time and temperature is then formulated in the framework of information geometry, in which the Fisher information metric plays an essential role. 
Usual recipe of the differential geometry in the two-dimensional parameter space of $\beta$ and $t$ leads to a simple equation for the scalar curvature as $R = -1$.
Then, we can make a reversed logic that this basic equation $R = -1$ has a solution for the Fisher information metric to characterize the relaxation process.
Investigating the global behaviors of the metric tensors, we see a thermodynamic emergence of time. 

\par

This study has focused on a special case of the OM process for the nonequilibrium relaxation phenomena. 
Therefore, we do not know how large family of nonequilibrium processes can be described in terms of the present simple relation of $R = -1$ obtained for the OM process. 
More generally, we may expect some equations such as dimensionless $R = $ constant to describe larger family of dynamical processes.

\par 

We have thus found that the time and the temperature regulate and generate each other. 
In particular, the present analysis provides a simple model to substantiate a concept that the time as a coarse-grained parameter appears in thermal manner 
through measurement or occurrence of events.

\section*{Acknowledgements}

\noindent
The author would like to acknowledge the Grants-in-Aid for Scientific Research (Nos.\ 17H06353 and 18K03825) from the Ministry of Education, Cultute, Sports, Science and Technology (MEXT).


\newpage


\section*{Figure captions}

\noindent
Figure 1: Plots of the likelihood $P(\{x_i\}|\lambda) = \prod_{i}P(x_i|\lambda)$ in Eq.\ (7) when we use Eq.\ (5) with $k, D, x_{0} = 1$ and $\{x_i\} = \{0.25,0.30,0.35,0.37,0.40\}$. 
(a) $\lambda = \beta$ and $t = 0.5$. (b) $\lambda = t$ and $\beta = 5.0$. 

\vspace{0.5cm}

\noindent
Figure 2: (a) Fisher information metric $g_{\beta\beta}$ as a function of $t$ for $\beta = 0.1, 1, 10$. 
(b) Fisher information metric $g_{\beta\beta}$ as a function of $\beta$ for $t = 0.1, 1, 10$. 
Model parameters have been employed as $k, D, x_{0} = 1$ for illustration.

\vspace{0.5cm}

\noindent
Figure 3: (a) Fisher information metric $g_{tt}$ as a function of $t$ for $\beta = 0.1, 1, 10$. 
(b) Fisher information metric $g_{tt}$ as a function of $\beta$ for $t = 0.1, 1, 10$. 
Model parameters have been employed as $k, D, x_{0} = 1$ for illustration.

\vspace{0.5cm}

\noindent
Figure 4: (a) Fisher information metric $g_{\beta t}$ as a function of $t$ for $\beta = 0.1, 1, 10$. 
(b) Fisher information metric $g_{\beta t}$ as a function of $\beta$ for $t = 0.1, 1, 10$. 
Model parameters have been employed as $k, D, x_{0} = 1$ for illustration.



\end{document}